\documentclass[12pt,thmsa]{article}
\usepackage{amssymb}

\usepackage{sw20lart}

\input{tcilatex}
\begin{document}

\author{Emilio Santos}
\title{Interpretation of Feynman\'{}s formalism of quantum mechanics in terms of
probabilities of paths }
\maketitle

\begin{abstract}
Feynman\'{}s path integrals formalism for non-relativistic quantum mechanics
is revisited. A comparison is made with the cases of light progagation
(Huygens principle) and Brownian motion. The difficulties for a physical
model behind Feynman\'{}s formalism are pointed out. It is proposed a
reformulation where the transition probability from one space-time point to
another one is the sum of probabilities of the possible paths. The Born
approximation for scattering is derived within the formalism, which suggests
an interpretation in terms of particles, without the need of Born\'{}s
assumption that the modulus squared of the wavefunction is a probability
density.

\end{abstract}

\section{Feynman formulation of quantum mechanics}

The path integral formulation of quantum mechanics was introduced by Feynman
in 1948\cite{Feynman}. In his book on the subject\cite{FeynmanHibbs} the
formalism is shown to be a straightforward consequence of the superposition
principle. In fact, let us assume that there is a source of particles at
point $x=x_{0},y=0$, a screen with two slits at $x=\pm a,y=b$, and the
particles may be detected at any point in the plane $y=c$. (For simplicity I
ignore here the third coordinate, $z$). If a particle leaves the source at
time $t=0$, crosses a slit at $t_{1}$ and arrives at the detector at $t_{2}$%
, the probability, $P(x),$ of reaching a point with coordinate $x$ in the
screen $y=c$ is proportional to the modulus squared of a ``probability
amplitude'', and the latter is the sum of two amplitudes, that is 
\begin{eqnarray}
P(x) &\varpropto &\left| A(x_{0},0\mid x,t_{2})\right| ^{2},\smallskip
A(x_{0},0\mid x,t_{2})=  \label{Feyn1} \\
&=&A(x_{0},0\mid a,t_{1})A(a,t_{1}\mid x,t_{2})+A(x_{0},0\mid
-a,t_{1})A(-a,t_{1}\mid x,t_{2}).  \nonumber
\end{eqnarray}
Now we consider the case of having many screens at positions $%
y_{1},y_{2},... $, where the particle may arrive at times $t_{1},t_{2},...$,
respectively, and that every screen possesses many slits, at positions $%
x_{1},x_{2}$, ...In this case the amplitude eq.$\left( \ref{Feyn1}\right) $
is replaced by 
\begin{equation}
A(x_{0},0\mid x,t)=\sum_{j}\sum_{k}...\sum_{q}A(x_{0},0\mid
x_{j},t_{1})A(x_{j},t_{1}\mid x_{k},t_{2})...A(x_{q},t_{n}\mid x,t).
\label{Feyn5}
\end{equation}
Here the set of positions $\left\{ x_{0},x_{j},x_{k},...x_{q},x\right\} $
may be called a path, so that the amplitude $A(x_{0},0\mid x,t)$ is a sum of
amplitudes, every one corresponding to one possible path. If the number of
slits in every screen increases indefinitely, at the end there will be no
screen at all. Then the discrete values $x_{l},x_{k},...$become continuous
and the sums become integrals, that is 
\begin{equation}
A(x_{0},0\mid x,t)=\int dx_{1}...\int dx_{n-1}A(x_{0},0\mid
x_{1},t_{1})...A(x_{n-1},t_{n-1}\mid x,t).  \label{Feyn6}
\end{equation}
The time intervals may be chosen identical, that is $t_{j+1}-t_{j}=%
\varepsilon ,$ with $\varepsilon $ as small as desired. In the limit $%
\varepsilon \rightarrow 0,$ $A(x_{0},0\mid x,t)$ becomes an integral of path
amplitudes$.$

The derivation shows that eq.$\left( \ref{Feyn6}\right) $ is not specific of
Feynman's formulation of quantum mechanics, but it may be valid for any
field fulfilling a superposition property, e. g. classical optics. In this
case, however, $P(x)$ of eq.$\left( \ref{Feyn1}\right) $ would not be a
probability but a light intensity, as will be illustrated below with
Huygen\'{}s principle. Also, an expression formally similar to eq.$\left( 
\ref{Feyn5}\right) $ may be used in the study of random motion when the
probability that a particle goes from one position to another one in some
time interval is the sum of probabilities corresponding to different paths,
a method pioneered by Norbert Wiener in the study of Brownian motion. This
will be illustrated below with a derivation of the diffusion law via a path
integral. In this case the probability density is given directly, rather
than via the modulus squared of an amplitude. More generally, a path
integrals formulation might be associated to the evolution of any physical
system if it is governed by a linear partial differential equation.

What is specific of Feynman\'{}s formulation of (non-relativistic) quantum
mechanics is the choice of $A$ to be the exponential of $i/ 
\rlap{\protect\rule[1.1ex]{.325em}{.1ex}}h%
$ times the (classical) Lagrangian, $L$, of the particle\'{}s motion, an
idea taken from Dirac\cite{Dirac}. For instance in the case of
one-dimensional motion in a potential $V(x)$ we have $L=\frac{1}{2}m%
\stackrel{\cdot }{x}^{2}-V(x)$ whence 
\begin{eqnarray}
A\left( x_{j-1},t_{j-1}\mid x_{j},t_{j}\right) &=&\sqrt{\frac{m}{2\pi i 
\rlap{\protect\rule[1.1ex]{.325em}{.1ex}}h%
\varepsilon }}\exp (\frac{i\varepsilon }{
\rlap{\protect\rule[1.1ex]{.325em}{.1ex}}h%
}L),  \label{Feyn4} \\
L &\equiv &\frac{1}{2}m\left( \frac{x_{j}-x_{j-1}}{\varepsilon }\right) ^{2}-%
\frac{1}{2}\left[ V\left( x_{j-1}\right) +V\left( x_{j}\right) \right] 
\nonumber
\end{eqnarray}
where $m$ is the mass of the particle. (This expression differs from the
original one of Feynman\cite{FeynmanHibbs} because I have substituted $\frac{%
1}{2}\left[ V\left( x_{j-1}\right) +V\left( x_{j}\right) \right] $ for $%
V\left[ \left( x_{j-1}+x_{j}\right) /2\right] $ for later convenience. Also
I ignore the terms $V\left( x_{0}\right) $ and $V\left( x\right) ,$ so that
the potential enters $n-1$ times in eq.$\left( \ref{Feyn6}\right) ,$ as it
should. Both formulations agree in the limit $\varepsilon \rightarrow 0).$
It is common to take the continuous limit and write eq.$\left( \ref{Feyn5}%
\right) $ in the form 
\begin{eqnarray}
A(x_{0},0 &\mid &x,t)=\int \mathcal{D}(paths)\exp \left[ \frac{i}{%
\rlap{\protect\rule[1.1ex]{.325em}{.1ex}}h%
}\int_{0}^{t}dtL(x,\stackrel{\cdot }{x})\right]  \nonumber \\
&=&\int \mathcal{D}(paths)\exp \left[ \frac{i}{%
\rlap{\protect\rule[1.1ex]{.325em}{.1ex}}h%
}\int_{0}^{t}dt\left( \frac{1}{2}m\stackrel{\cdot }{x}^{2}-V(x)\right)
\right] .  \label{Feyn3}
\end{eqnarray}
This symbolic equation actually means the limit $\varepsilon \rightarrow 0$
of eq.$\left( \ref{Feyn6}\right) $ with $A$ given by eq.$\left( \ref{Feyn4}%
\right) .$ The amplitude eq.$\left( \ref{Feyn3}\right) $ may be calculated
explicitly for simple potentials\cite{FeynmanHibbs}.

The amplitude $A(x_{0},0\mid x,t)$ is named the ``propagator''. The
interesting property is that the propagator allows getting the wavefunction
at time $t$ from the wavefunction at time $0$, that is\cite{FeynmanHibbs} 
\begin{equation}
\psi \left( x,t\right) =\int dx_{0}\psi \left( x_{0},t\right) A(x_{0},0\mid
x,t).  \label{F9}
\end{equation}
Hence it follows that the propagator fulfils the Schr\"{o}dinger equation
with the initial condition 
\[
A(x_{0},0\mid x,0)=\delta \left( x-x_{0}\right) , 
\]
where $\delta \left( x\right) $ is Dirac\'{}s delta. Thus the propagator is
the Green\'{}s function of the Schr\"{o}dinger equation.

As is well known the path integrals formulation may be generalized to 3
dimensions, to many-particles quantum mechanics and also to relativistic
field theory. It has an extremely important role in modern theoretical
physics, both because it is well adapted to derive general properties, e. g.
symmetries, and due to the relevance for actual calculations, as shown by
the use of Feynman graphs in covariant perturbation theory\cite{ZinnJustin}.
However dealing with formal and calculational aspects lies out of the scope
of this article, which is devoted to the physical interpretation of the
Feynman formalism in (non-relativistic) quantum mechanics. In particular I
will present below a possible interpretation of the path integrals
formalism, but before I will revisit the use of a formula similar to eq.$%
\left( \ref{Feyn6}\right) $ in classical optics and diffusion theory in
order to make a comparison with the Feynman formalism.

\section{Huygens principle}

Historically the first formulation of a physical theory in terms of ``path
integrals'' goes back to Christiaan Huygens, more than three centuries ago.
In fact Huygens proposed that light are waves, in opposition to his
contemporary Isaac Newton who supported a corpuscular theory. He was able to
explain the straight line propagation and other properties of light from his
celebrated principle. Huygens principle states that light propagation may be
understood as if every point where light arrives becomes the source of a
spherical wave, and the waves coming from different points are able to
interfere. In practice this implies that from each point of a given
wavefront at time $t$ spherical wavelets originate. Thus Huygens principle
may be formalized stating that the light arriving at time $t$ at a point $%
\mathbf{r}$ may be calculated from the three-dimensional generalization of
eq.$\left( \ref{Feyn6}\right) $ with appropriate transition amplitudes $%
A\left( \mathbf{r}_{j},t_{j}\mid \mathbf{r}_{j+1},t_{j+1}\right) .$ At a
difference with the cases of Feynman\'{}s formalism, eq.$\left( \ref{Feyn3}%
\right) ,$ or diffusion (see below eq.$\left( \ref{dif1}\right) )$, where
the velocities of the particles may have any value, light travels in vacuum
with a fixed velicity, $c$, which puts the constraint $\left| \mathbf{r}%
_{j+1}-\mathbf{r}_{j}\right| =c\left( t_{j+1}-t_{j}\right) \mathbf{.}$ Hence
the transition amplitude should be of the form 
\begin{equation}
A\left( \mathbf{r}_{j},t_{j}\mid \mathbf{r}_{j+1},t_{j+1}\right) =f\left( 
\mathbf{r}_{j}\mid \mathbf{r}_{j+1}\right) \delta \left( \left| \mathbf{r}%
_{j+1}-\mathbf{r}_{j}\right| -c\left( t_{j+1}-t_{j}\right) \right) .
\label{Huyg1}
\end{equation}
Light consists of transverse waves (that is the electric and magnetic fields
are perpendicular to the direction of propagation) which implies that the
function $f$ should depend also on the angle between the electric field and
the $\mathbf{r}_{j}-\mathbf{r}_{j+1}$ vectors. I shall avoid this
complication, irrelevant for our purposes, considering in the following
longitudinal waves, e. g. sound propagation in air.

For monocromatic sound waves in air the function $f$ of eq.$\left( \ref
{Huyg1}\right) $ is specially simple, namely

\begin{equation}
f\left( \mathbf{r}_{j}\mid \mathbf{r}_{j+1}\right) =\frac{\exp (ik\left| 
\mathbf{r}_{j+1}-\mathbf{r}_{j}\right| )}{2\pi i\left| \mathbf{r}_{j+1}-%
\mathbf{r}_{j}\right| }.  \label{Huyg2}
\end{equation}
where the denominator takes into account that the intensity from a point
source decreases as the inverse of the distance squared. I stress that here
the use of complex amplitudes, i. e. the introduction of the imaginary unit
number $i$, has no deep meaning, it is just a convenient mathemtical
procedure to simplify the calculations. Actually the wave amplitudes might
be always represented by real numbers (indeed the light amplitude may be
taken to be the electric field of the radiation and the amplitude of sound
vaves in air to be the excess pressure).

In practice it is most frequent to use very simple ``paths'' consisting of
just three points, namely \{$\mathbf{r}_{1},\mathbf{r}_{2},\mathbf{r}$\}.
For instance a typical problem solved with that choice is the difraction by
a small hole (see any book of electromagnetic theory, e.g. \cite{Jackson}).

\section{Diffusion}

The path integral formulation is most intuitive in the case of diffusion.
Let us consider the well known problem of the random walk in one dimension.
It consists of a particle that travels a distance $\lambda $, either towards
the left or towards the right with equal probability, the step taking a time 
$\varepsilon $. During the next time interval $\varepsilon $ the particle
travels again a distance $\lambda $ either to the left or to the right with
equal probability, and so on. The problem is to find the probability that
the particle is at a given distance, say $x=l\lambda $, of the origin after $%
n$ steps, $n$ being a natural number and $l$ an integer, $\left| l\right| $ $%
\leq n.$ The problem might be stated by means of a sum of paths as follows 
\begin{equation}
P(x_{0},0\mid x,t)=\sum_{j}...\sum_{q}P(x_{0},0\mid x_{j},\varepsilon
)P(x_{j},\varepsilon \mid x_{k},2\varepsilon )...P(x_{q},(n-1)\varepsilon
\mid x,t),  \label{dif0}
\end{equation}
where the set of positions $\left\{ x_{0},x_{j},x_{k},...x_{q},x\right\} ,$
represents a path. Remember that $\left| x_{j}-x_{0}\right| =\left|
x_{k}-x_{j}\right| =...=\lambda $ and that the probabilies for every step
are 
\begin{equation}
P(x_{s},m\varepsilon \mid x_{r},(m+1)\varepsilon )=\frac{1}{2}\text{ if }%
x_{r}-x_{s}=\lambda ,P=\frac{1}{2}\text{ if }x_{r}-x_{s}=-\lambda .
\label{dif00}
\end{equation}
The solution may be found as follows. In order that the particle arrives at
a distance $l\lambda $ from the origin after $n$ steps, it is necessary and
sufficient that $k\equiv (n+l)/2$ steps are towards the right and $%
n-k=(n-l)/2$ towards the left (assuming that the positive direction is to
the right.) The set of positions may be called \emph{the path} of the
particle. The probability of a given path is $\left( 1/2\right) ^{n}$ and
the number of paths leading from $0$ to $l\lambda $ is the combinatorial
number $\binom{n}{{}{{}{{}{{}{{}{{}{}}}}k}}}$ As a consequence the
probability of reaching $l\lambda $ in $n$ steps is 
\[
{}{{}{{}{{}{{}{{}{{P_{mL}=}\binom{n}{{}{{}{{}{{}{{}{{}{}}}}k}}}{\left( \frac{%
1}{2}\right) ^{n}=\frac{n!}{k!\left( n-k\right) !}\left( \frac{1}{2}\right)
^{n}=\frac{n!}{\left[ (n+l)/2\right] !\left[ \left( (n-l)/2\right) \right] !}%
\left( \frac{1}{2}\right) ^{n}.}}}}}}} 
\]
If $n$ and $m$ are very large, we may approximate the factorials by the
Stirling formula, that is 
\[
\log \left( k!\right) \simeq \frac{1}{2}\log \left( 2\pi k\right) +k\log
k-k. 
\]
and we get, ignoring terms which do not depend on $l$, 
\[
\log P_{x}\simeq const.-\frac{n+l}{2}\log \left( \frac{n+l}{2}\right) -\frac{%
n-l}{2}\log \left( \frac{n-l}{2}\right) . 
\]
Hence I obtain 
\[
P_{x}\varpropto \left( n+l\right) ^{\frac{n+l}{2}}(n-l)^{\frac{n-l}{2}%
}\varpropto \left( 1-\frac{l^{2}}{n^{2}}\right) ^{n/2}\left( \frac{n-l}{n+l}%
\right) ^{m/2}\simeq \exp \left( -\frac{l^{2}}{2n}\right) , 
\]
where I have taken into account that in most paths $l<<n$ for very large $n$%
. In terms of the total time, $t\equiv n\varepsilon ,$ and the final
position $x\equiv n\lambda $, this may be written as a probability density
which properly normalized reads 
\begin{equation}
\rho \left( x_{0},0\mid x,t\right) =\sqrt{\frac{1}{4\pi t}}\exp \left( -%
\frac{x^{2}}{4Dt}\right) ,  \label{dif2}
\end{equation}
the parameter $D\equiv \lambda ^{2}/\varepsilon $ being the diffusion
constant. For us the relevant conclusion is that the diffusion probability
may be calculated as a sum of the probabilities corresponting to all paths
leaving the origin at the initial time and arriving at the point $x$ at time 
$t$. The calculation is rather simple because all paths have the same
probability.

We may also derive the diffusion probability assuming that in every step of
duration $\varepsilon $ the particle may travel any distance, $\Delta x,$
with a probability proportional to $\exp \left( -\frac{\Delta x^{2}}{%
2D\varepsilon }\right) .$ Then the probability of reaching $x$ will be 
\begin{equation}
P\varpropto \int dx_{1}\int dx_{2}...\int dx_{n-1}\exp \left[ -\frac{1}{4D}%
\sum_{k=0}^{n-1}\frac{\left( x_{k+1}-x_{k}\right) ^{2}}{t_{m+1}-t_{m}}%
\right] ,x_{n}=x,  \label{dif3}
\end{equation}
In the limit of large $n$, but fixed $t=n\varepsilon $, this may be written
substituting integrals for the sums, that is 
\begin{equation}
P\varpropto \int \mathcal{D}(paths)\exp \left( \frac{1}{4D}\int_{0}^{t}dt%
\stackrel{\cdot }{x}^{2}\right) ,  \label{dif1}
\end{equation}
which is formally similar to the Feynman path integral eq.$\left( \ref{Feyn3}%
\right) $. There are however two fundamental differences. Firstly in the
diffusion case eq.$\left( \ref{dif1}\right) $ gives directly the
probability, rather than the amplitude, and secondly the quantity inside the
exponential is real, so that the sum involves positive probabilities, rather
than complex numbers. The integrals in eq.$\left( \ref{dif3}\right) $ are
rather simple and the result is the probability density eq.$\left( \ref{dif2}%
\right) ,$ which is the Green\'{}s function of the diffusion equation 
\[
\frac{\partial }{\partial t}\rho (0,0\mid x,t)=D\frac{\partial ^{2}}{%
\partial x^{2}}\rho (0,0\mid x,t). 
\]

A technique similar to the one illustrated here for diffusion may be applied
to any process where the probability of going from an initial state to a
final one is the sum of the probabilities of the different paths. This is
the case for any random motion, but the fact that the transition probability
in a step does not depend on the previous positions is not general. If it is
fulfilled the stochastic process is called Markovian. This property holds
true in eq.$\left( \ref{dif0}\right) $ where the transition probabilities in
different steps are statistically indpendent of each other.

\section{Physical interpretation of path integrals in quantum mechanics}

It is sometimes stated that Feynman\'{}s path integral formalism provides an
intuitive picture of the quantum mechanical evolution. I do not agree. The
formalism of non-relativistic quantum mechanics is counterintuitive for two
reasons. Firstly because Feynman paths are just sets of disconected points,
rather than (continuous) paths in the usual sense of the word. Secondly
because in a intuitive picture the probability of travel of a particle
between two points should involve a sum of probabilities not amplitudes. I
comment on these two shortcomings in the following.

The alleged path represented by the succesive positions $%
\{x_{0},x_{1},x_{2}...x_{n},x\}$ is actually a set of points, each one
separated from the previous one by a long distance. Indeed the position $%
x_{j-1}$ may have values in the interval $\left( -\infty ,\infty \right) ,$
which is the interval of integration, all values having the same weight
according to the Riemann measure of the integral (see eq.$\left( \ref{Feyn6}%
\right) ).$ Similarly the position $x_{j}$ may have values in the interval $%
\left( -\infty ,\infty \right) $ all values having the same weight. Thus the
step $u_{j}=x_{j}-x_{j-1}$ between two positions may be \textit{arbitrarily
large.} Indeed its mean squared value, $\left\langle u_{j}^{2}\right\rangle
, $ diverges if calculated with the Riemann measure. The counterintuitive
character is enhanced by the fact that the (indefinitely long) step $u_{j}$%
\textit{\ takes place in an infinitesimal time interval }$\varepsilon .$
Thus Feynman\'{}s eq.$\left( \ref{Feyn6}\right) $ should be seen as a purely
mathematical construction, an useful calculational tool where the physical
meaning appears only in the final result of the calculation.

In sharp contrast Huygens\'{} path integrals, eqs.$\left( \ref{Huyg1}\right)
,$ are continuous$.$ In fact the quantities$\left| \mathbf{r}_{j+1}-\mathbf{r%
}_{j}\right| $ are never too large due to the denominator in eq.$\left( \ref
{Huyg2}\right) $ and, above all, they decrease to zero when the time
interval, $t_{j+1}-t_{j},$ goes to zero. Thus the sum or integral involved
is over \emph{continuous paths}. A similar thing happens in the diffusion
problem defined by eqs.$\left( \ref{dif0}\right) $ and $\left( \ref{dif00}%
\right) .$ In this case every path is continuous although
non-differentiable. The same is true if we define the path via eq.$\left( 
\ref{dif3}\right) $ where the probability that the mean velocity in a step, $%
(x_{k+1}-x_{k})/(t_{m+1}-t_{m}),$ surpasses a value $K$ goes exponentially
to zero when $K\rightarrow \infty .$

From a physical rather than formal point of view, Feynman\'{}s path integral
is more similar to Huygen\'{}s principle of classical optics than to the
diffusion problem. Indeed the use of an amplitude suggests a wave picture
although the fact that the modulus squared of the amplitude is a
probability, rather than an intensity, gives a particle appearance. However
the particle picture is misleading. In fact, in spite of the frequent use of
the expression ``probability amplitude'' in quantum theory, I am unable to
give a real meaning to these two words. In contrast the interpretation of
Feynman\'{}s path integrals for radiation (``photons'') is more intuitive.
It might be considered as an elaboration of Huygens principle of classical
optics. As said above the main conceptual difference is that in classical
optics the amplitude squared is an intensity (power per unit area) whilst in
Feynman\'{}s formalism it is assumed a probability density (of having a
photon at a given point of space-time). But there is no real problem in
assuming also in the latter formalism the intensity interpretation if we
suppose that, at the time of detection, the probability is proportional to
the intensity arriving at the detector, with some peculiarities\cite{SO}.

\section{Transition probability as a sum of probabilities of the paths}

In the following I propose a possible intuitive interpretation of
Feynman\'{}s formalism in the case of particles. It leads to the probability
as a sum of path\'{}s probabilities\cite{Santos73}. This method might
provide a physical model of the motion of quantum particles, becoming an
alternative to the popular de Broglie-Bohm theory (or Bohmian mechanics).

I start considering an integral of pairs of paths, starting from eq.$\left( 
\ref{Feyn3}\right) ,$ as follows 
\begin{eqnarray}
P(x_{0},0 &\mid &x,t)\varpropto \left| A(x_{0},0\mid x,t)\right| ^{2} 
\nonumber \\
&=&\int \mathcal{D}(pathpairs)\exp \left[ \frac{i}{%
\rlap{\protect\rule[1.1ex]{.325em}{.1ex}}h%
}\int_{0}^{t}dt\left( \frac{1}{2}m\stackrel{\cdot }{x}^{2}-V(x)\right)
\right]  \nonumber \\
&&\times \exp \left[ -\frac{i}{
\rlap{\protect\rule[1.1ex]{.325em}{.1ex}}h%
}\int_{0}^{t}dt\left( \frac{1}{2}m\stackrel{\cdot }{y}^{2}-V(y)\right)
\right] .  \label{pairs}
\end{eqnarray}
After that I introduce the new variables $x+y\equiv 2z,x-y\equiv w$ and
integrate over $w$, thus getting an integral over paths, every one in terms
of the variable $z$. The interpretation of the quantum motion as a kind of
random motion would be possible if the probability of every path is never
negative. However a general proof that this is the case will not be made in
this paper ( see arXiv: quant-ph, 1603.02215), but I shall study only a few
particular cases where the positivity holds true. Firslty I shall consider
the case of free propagation, that is $V=0$. Thus I get 
\begin{eqnarray*}
P(x_{0},0 &\mid &x,t)\varpropto \lim_{\varepsilon \rightarrow 0}\left( \frac{%
m}{2\pi 
\rlap{\protect\rule[1.1ex]{.325em}{.1ex}}h%
\varepsilon }\right) ^{n}\int dx_{1}...\int dx_{n-1}\int dy_{1}...\int
dy_{n-1} \\
&&\times \prod_{j=1}^{n}\exp \left[ \frac{im}{2 
\rlap{\protect\rule[1.1ex]{.325em}{.1ex}}h%
\varepsilon }\left( x_{j}-x_{j-1}\right) ^{2}\right] \exp \left[ -\frac{im}{%
2 
\rlap{\protect\rule[1.1ex]{.325em}{.1ex}}h%
\varepsilon }\left( y_{j}-y_{j-1}\right) ^{2}\right] .
\end{eqnarray*}
With the change of variables above mentioned, this becomes after some
algebra 
\begin{eqnarray}
P(x_{0},0 &\mid &x,t)\varpropto \lim_{\varepsilon \rightarrow 0}\left( \frac{%
m}{2\pi 
\rlap{\protect\rule[1.1ex]{.325em}{.1ex}}h%
\varepsilon }\right) ^{n}\int dz_{1}...\int dz_{n-1}\int dw_{1}...\int
dw_{n-1}  \nonumber \\
&&\times \prod_{j=1}^{n-1}\exp \left[ \frac{im}{%
\rlap{\protect\rule[1.1ex]{.325em}{.1ex}}h%
\varepsilon }w_{j}\left( z_{j-1}-2z_{j}+z_{j+1}\right) \right] ,
\label{Feyn14}
\end{eqnarray}
where I have taken into account that $w_{0}=w_{n}=0.$ Hence, as $%
z_{0}=x_{0},z_{n}=x,$ I get 
\begin{equation}
P(x_{0},0\mid x,t)\varpropto \lim_{\varepsilon \rightarrow 0}\frac{m}{2\pi 
\rlap{\protect\rule[1.1ex]{.325em}{.1ex}}h%
\varepsilon }\int dz_{1}...\int dz_{n-1}\prod_{j=1}^{n-1}\delta \left(
z_{j+1}-2z_{j}+z_{j-1}\right) .  \label{Feyn13}
\end{equation}
If the integrals in the variables $z_{j}$ are performed it is easy to check
that the result agrees with the standard one\cite{FeynmanHibbs}$.$ Eq.$%
\left( \ref{Feyn13}\right) $ should be interpreted as a propagation with
constant velocity, $v=\left( x-x_{0}\right) /t.$ In fact the Dirac deltas $%
\delta \left( z_{j+1}-2z_{j}+z_{j-1}\right) $ imply ($%
z_{j+1}-2z_{j}+z_{j-1})/\varepsilon ^{2}=0$, that is nil acceleration.

Now I will generalize the procedure to three dimensions and to the motion in
the presence of a potential $V(\mathbf{r})$. In this case, steps similar to
those leading to eq.$\left( \ref{Feyn4}\right) $ give

\begin{eqnarray}
P(\mathbf{r}_{0},0 &\mid &\mathbf{r},t)\varpropto \lim_{\varepsilon
\rightarrow 0}\frac{m}{2\pi 
\rlap{\protect\rule[1.1ex]{.325em}{.1ex}}h%
\varepsilon }\left( 2\pi \right) ^{3-3n}\int d\mathbf{r}_{1}...\int d\mathbf{%
r}_{n-1}\int d\mathbf{v}_{1}...\int d\mathbf{v}_{n-1}  \nonumber \\
\, &&\times \prod_{j=1}^{n-1}\exp \left[ i\mathbf{v}_{j}\cdot \mathbf{s}_{j}+%
\frac{i\varepsilon }{
\rlap{\protect\rule[1.1ex]{.325em}{.1ex}}h%
}V\left( \mathbf{r}_{j}-\frac{
\rlap{\protect\rule[1.1ex]{.325em}{.1ex}}h%
\varepsilon \mathbf{v}_{j}}{m}\right) -\frac{i\varepsilon }{%
\rlap{\protect\rule[1.1ex]{.325em}{.1ex}}h%
}V\left( \mathbf{r}_{j}+\frac{
\rlap{\protect\rule[1.1ex]{.325em}{.1ex}}h%
\varepsilon \mathbf{v}_{j}}{m}\right) \right] ,  \label{F15}
\end{eqnarray}
where 
\[
\mathbf{v}_{j}\equiv \frac{m}{2
\rlap{\protect\rule[1.1ex]{.325em}{.1ex}}h%
\varepsilon }\mathbf{w}_{j},\mathbf{s}_{j}\equiv \mathbf{r}_{j-1}-2\mathbf{r}%
_{j}+\mathbf{r}_{j+1} 
\]

Performing the integrals in $\mathbf{v}_{j}$ is not possible without a
knowledge of potential, $V\left( \mathbf{r}\right) ,$ but to lowest order in 
$
\rlap{\protect\rule[1.1ex]{.325em}{.1ex}}h%
$ it is simple. It is obtained 
\begin{equation}
P(\mathbf{r}_{0},0\mid \mathbf{r},t)\varpropto \lim_{\varepsilon \rightarrow
0}\frac{m}{2\pi 
\rlap{\protect\rule[1.1ex]{.325em}{.1ex}}h%
\varepsilon }\int d\mathbf{r}_{1}...\int d\mathbf{r}_{n-1}\prod_{j=1}^{n-1}%
\delta ^{3}\left( \mathbf{s}_{j}+\frac{2\varepsilon ^{2}}{m}\mathbf{\nabla }%
V(\mathbf{r}_{j})\right) ,  \label{F0}
\end{equation}
which corresponds to a motion fulfilling at every time 
\[
m\frac{\mathbf{r}_{j-1}-2\mathbf{r}_{j}+\mathbf{r}_{j+1}}{\varepsilon ^{2}}=-%
\mathbf{\nabla }V(\mathbf{r}_{j}), 
\]
that is the classical equation of motion. Thus we get the classical limit of
quantum mechanics.

It is interesting that the same result, eq.$\left( \ref{F0}\right) ,$ is
obtained if the potential, $V(\mathbf{r}_{j}),$ is at most quadratic in the
coordinates, which should be interpreted saying that in linear problems the
quantum particle follows the classical path. The typical example is the
harmonic oscillator. This is the reason why quantum mechanics of linear
systems looks semiclassical. In this case all quantum effects come from the
fact that the initial wave function cannot be localized in a too small
region due to the Heisenberg uncertainty principle, a constraint which does
not appear directly in our path integrals formalism. We see that some
information is lost in passing from the (Feynman\'{}s) sum of amplitudes of
paths to our sum of probabilities of paths, like eq.$\left( \ref{F0}\right)
. $

\section{The Born approximation}

With the formalism here presented it is easy to derive Born\'{}s
approximation for scattering as follows. I consider an experiment where a
particle leaves a source placed at position $\mathbf{r}_{0}$ at time $t=0$,
it moves freely (i. e. in the absence of forces) until about time $t_{k},$
where the particle enters a region with a potential. Then the particle is
deflected by the potential $V\left( \mathbf{r}\right) $ and, after leaving
the interaction region, it may move freely towards a detector placed at the
point $\mathbf{r}.$ In order to calculate the probability of arrival I will
start from eq.$\left( \ref{F15}\right) $ and make a calculation involving
some approximations resting upon the fact that the potential is weak in some
sense to be specified, this being the basic hypothesis of the Born
approximation. Firstly I restrict the potential to act on the particle only
once, at time $t_{k}$. Thus we get from eq.$\left( \ref{F15}\right) $%
\begin{eqnarray}
P &\varpropto &\lim_{\varepsilon \rightarrow 0}\varepsilon ^{-1}\int d%
\mathbf{r}_{1}...\int d\mathbf{r}_{n-1}\int d\mathbf{v}_{1}...\int d\mathbf{v%
}_{n-1}\prod_{j=1}^{k-1}\exp \left( i\mathbf{v}_{j}\cdot \mathbf{s}%
_{j}\right)  \label{F7} \\
&&\times \exp \left[ i\varepsilon \mathbf{v}_{k}\cdot \mathbf{s}_{k}+\frac{%
i\varepsilon }{
\rlap{\protect\rule[1.1ex]{.325em}{.1ex}}h%
}V\left( \mathbf{r}_{k}-\frac{
\rlap{\protect\rule[1.1ex]{.325em}{.1ex}}h%
\varepsilon \mathbf{v}_{k}}{m}\right) -\frac{i\varepsilon }{%
\rlap{\protect\rule[1.1ex]{.325em}{.1ex}}h%
}V\left( \mathbf{r}_{k}+\frac{
\rlap{\protect\rule[1.1ex]{.325em}{.1ex}}h%
\varepsilon \mathbf{v}_{k}}{m}\right) \right] \prod_{l=k+1}^{n-1}\exp \left(
i\mathbf{v}_{l}\cdot \mathbf{s}_{l}\right) ,  \nonumber
\end{eqnarray}
where I have removed a constant factor, irrelevant because only the relative
probability is calculated (hence the use of the proportionality simbol, $%
\varpropto .)$ For later convenience I will write the exponential involving $%
\mathbf{v}_{k}$ in the form 
\[
\exp \left( i\varepsilon \mathbf{v}_{k}\cdot \mathbf{s}_{k}\right) +\exp
\left( i\varepsilon \mathbf{v}_{k}\cdot \mathbf{s}_{k}\right) \left\{ \exp
\left[ \frac{i\varepsilon }{
\rlap{\protect\rule[1.1ex]{.325em}{.1ex}}h%
}V\left( \mathbf{r}_{k}-\frac{
\rlap{\protect\rule[1.1ex]{.325em}{.1ex}}h%
\varepsilon \mathbf{v}_{k}}{m}\right) -\frac{i\varepsilon }{%
\rlap{\protect\rule[1.1ex]{.325em}{.1ex}}h%
}V\left( \mathbf{r}_{k}+\frac{
\rlap{\protect\rule[1.1ex]{.325em}{.1ex}}h%
\varepsilon \mathbf{v}_{k}}{m}\right) \right] -1\right\} , 
\]
so that eq.$\left( \ref{F7}\right) $ becomes 
\begin{eqnarray}
P &\varpropto &\lim_{\varepsilon \rightarrow 0}\varepsilon ^{-1}\int d%
\mathbf{r}_{1}...\int d\mathbf{r}_{n-1}\int d\mathbf{v}_{1}...\int d\mathbf{v%
}_{n-1}\prod_{j=1}^{n-1}\exp \left( i\mathbf{v}_{j}\cdot \mathbf{s}%
_{j}\right)  \label{F13} \\
&&+\lim_{\varepsilon \rightarrow 0}\varepsilon ^{-1}\int d\mathbf{r}%
_{1}...\int d\mathbf{r}_{n-1}\int d\mathbf{v}_{1}...\int d\mathbf{v}%
_{n-1}\prod_{j=1}^{k-1}\exp \left( i\mathbf{v}_{j}\cdot \mathbf{s}%
_{j}\right) \exp \left( i\varepsilon \mathbf{v}_{k}\cdot \mathbf{s}%
_{k}\right)  \nonumber \\
&&\times \left\{ \exp \left[ \frac{i\varepsilon }{%
\rlap{\protect\rule[1.1ex]{.325em}{.1ex}}h%
}V\left( \mathbf{r}_{k}-\frac{
\rlap{\protect\rule[1.1ex]{.325em}{.1ex}}h%
\varepsilon \mathbf{v}_{k}}{m}\right) -\frac{i\varepsilon }{%
\rlap{\protect\rule[1.1ex]{.325em}{.1ex}}h%
}V\left( \mathbf{r}_{k}+\frac{
\rlap{\protect\rule[1.1ex]{.325em}{.1ex}}h%
\varepsilon \mathbf{v}_{k}}{m}\right) \right] -1\right\}
\prod_{l=k+1}^{n-1}\exp \left( i\mathbf{v}_{l}\cdot \mathbf{s}_{l}\right) . 
\nonumber
\end{eqnarray}
The first term takes into account the possibility that the particle goes
from the source to the detector in a straight line, without crossing the
target. In the experimental practice this possibility is eliminated using
appropriate collimators, and I shall ignore this term.

In the following I shall study the second term of eq.$\left( \ref{F13}%
\right) .$ The integrals in all the variables $\mathbf{v}_{j}$ and $\mathbf{v%
}_{l}$ are trivial and give Dirac\'{}s deltas, which implies 
\[
\mathbf{s}_{j}\equiv \mathbf{r}_{j-1}-2\mathbf{r}_{j}+\mathbf{r}_{j+1}=0,%
\mathbf{s}_{j}\equiv \mathbf{r}_{l-1}-2\mathbf{r}_{l}+\mathbf{r}_{l+1}=0, 
\]
for all $j$ and $l$. This means that the particle moves in a straight line
with constant velocity, $\mathbf{v}_{in}$, from $\mathbf{r}_{0}$ to $\mathbf{%
r}_{k}$ and again with constant velocity, $\mathbf{v}_{out}$, from $\mathbf{r%
}_{k}$ to $\mathbf{r}.$ Thus we have 
\begin{equation}
\mathbf{v}_{in}\equiv \frac{\mathbf{r}_{k}-\mathbf{r}_{0}}{k\varepsilon }=%
\frac{\mathbf{r}_{k}-\mathbf{r}_{k-1}}{\varepsilon },\mathbf{v}_{out}\equiv 
\frac{\mathbf{r}-\mathbf{r}_{k}}{\left( n-k\right) \varepsilon }=\frac{%
\mathbf{r}_{k+1}-\mathbf{r}_{k}}{\varepsilon },  \label{F6}
\end{equation}
which implies 
\[
\mathbf{s}_{k}\equiv \mathbf{r}_{k-1}-2\mathbf{r}_{k}+\mathbf{r}%
_{k+1}=\varepsilon \left( \mathbf{v}_{out}-\mathbf{v}_{in}\right) \equiv
\varepsilon \Delta \mathbf{v.} 
\]
Our aim is to get the probability that the final velocity (in the detector)
is $\mathbf{v}_{out}$ for given initial velocity (in the source) $\mathbf{v}%
_{in}$. This will be the sum of probabilities of the possible paths with the
velocity change $\Delta \mathbf{v.}$ Each path is defined by the the initial
and final positions plus a point $\mathbf{r}_{k}$ in the target. However the
region where the potential $V\left( \mathbf{r}_{k}\right) $ is relevant (the
target) is microscopic, whilst the distances from source to target, $\left| 
\mathbf{r}_{k}-\mathbf{r}_{0}\right| ,$ and from target to detector, $\left| 
\mathbf{r}-\mathbf{r}_{k}\right| ,$ are both macroscopic. Hence $\Delta 
\mathbf{v}$ may be defined without a precise knowledge of the actual
positions of the particle in the source and the detector. Therefore an
accurate value of $\mathbf{r}_{0}$ and $\mathbf{r}$ are not required. The
integrals in $\mathbf{r}_{k}$ and $\mathbf{v}_{k}$, eq.$\left( \ref{F7}%
\right) ,$ that is 
\begin{equation}
\int d\mathbf{r}_{k}\int d\mathbf{v}_{k}\exp \left( i\varepsilon \mathbf{v}%
_{k}\cdot \mathbf{s}_{k}\right) \left\{ \exp \left[ \frac{i\varepsilon }{%
\rlap{\protect\rule[1.1ex]{.325em}{.1ex}}h%
}V\left( \mathbf{r}_{k}-\frac{
\rlap{\protect\rule[1.1ex]{.325em}{.1ex}}h%
\varepsilon \mathbf{v}_{k}}{m}\right) -\frac{i\varepsilon }{%
\rlap{\protect\rule[1.1ex]{.325em}{.1ex}}h%
}V\left( \mathbf{r}_{k}+\frac{
\rlap{\protect\rule[1.1ex]{.325em}{.1ex}}h%
\varepsilon \mathbf{v}_{k}}{m}\right) \right] -1\right\} ,  \label{F10}
\end{equation}
will be calculated as follows.

In order to simplify the notation, in the derivation I will use units $
\rlap{\protect\rule[1.1ex]{.325em}{.1ex}}h%
=\varepsilon =m=1$ and remove the subindices $k$ in eq.$\left( \ref{F10}%
\right) .$ The dimensional quantities will be restored at the end. As said
above the basis of the Born approximation is the assumption that the
potential is weak, which justifies to approximate the last factor of eq.$%
\left( \ref{F10}\right) $ by 
\begin{eqnarray}
&&\exp \left[ iV\left( \mathbf{r}-\mathbf{v}\right) -iV\left( \mathbf{r}+%
\mathbf{v}\right) \right] -1  \nonumber \\
&\simeq &iV\left( \mathbf{r}-\mathbf{v}\right) -iV\left( \mathbf{r}+\mathbf{v%
}\right) +\frac{1}{2}\left[ iV\left( \mathbf{r}-\mathbf{v}\right) -iV\left( 
\mathbf{r}+\mathbf{v}\right) \right] ^{2}  \nonumber \\
&\simeq &iV\left( \mathbf{r}-\mathbf{v}\right) -iV\left( \mathbf{r}+\mathbf{v%
}\right) -\frac{1}{2}\left[ V\left( \mathbf{r}-\mathbf{v}\right) \right]
^{2}-\frac{1}{2}\left[ V\left( \mathbf{r}+\mathbf{v}\right) \right] ^{2} 
\nonumber \\
&&+V\left( \mathbf{r}-\mathbf{v}\right) V\left( \mathbf{r}+\mathbf{v}\right)
.  \label{F14}
\end{eqnarray}
Now I will show that only the latter term gives a relevant contribution. In
fact we get 
\[
\int d\mathbf{r}\int d\mathbf{v}\exp \left( i\varepsilon \mathbf{v}\cdot
\Delta \mathbf{v}\right) V\left( \mathbf{r}-\mathbf{v}\right) =\int d\mathbf{%
v}\exp \left[ i\varepsilon \mathbf{v}\cdot \Delta \mathbf{v}\right] \int d%
\mathbf{x}V\left( \mathbf{x}\right) , 
\]
where I have introduced the new variable $\mathbf{x=r-v}$. Similarly with
the change $\mathbf{y=r+v}$ I obtain 
\[
\int d\mathbf{r}\int d\mathbf{v}\exp \left( i\varepsilon \mathbf{v}\cdot
\Delta \mathbf{v}\right) V\left( \mathbf{r}+\mathbf{v}\right) =\int d\mathbf{%
v}\exp \left[ i\varepsilon \mathbf{v}\cdot \Delta \mathbf{v}\right] \int d%
\mathbf{y}V\left( \mathbf{y}\right) . 
\]
We see that the contributions of the former two terms of eq.$\left( \ref{F14}%
\right) $ cancel out. The contribution of the next term is 
\[
-\frac{1}{2}\int d\mathbf{r}\int d\mathbf{v}\exp \left( i\varepsilon \mathbf{%
v}\cdot \Delta \mathbf{v}\right) \left[ V\left( \mathbf{r}-\mathbf{v}\right)
\right] ^{2}=\int d\mathbf{v}\exp \left[ i\varepsilon \mathbf{v}\cdot \Delta 
\mathbf{v}\right] \int d\mathbf{x}\left[ V\left( \mathbf{x}\right) \right]
^{2}, 
\]
and the integral in $\mathbf{v}$ is zero except if $\Delta \mathbf{v=0,}$
which corresponds to a motion in straight line from the source to the
detector, and the same is true for the contribution of the term with $\left[
V\left( \mathbf{r}+\mathbf{v}\right) \right] ^{2}.$ I shall ignore these
terms for the same reasons why I ignored the first term in eq.$\left( \ref
{F13}\right) .$ I conclude that in the expansion of the exponential eq.$%
\left( \ref{F14}\right) $ the term giving the leading contribution is the
following 
\begin{eqnarray}
P &\varpropto &\int d\mathbf{r}\int d\mathbf{v}\exp \left( i\varepsilon 
\mathbf{v}\cdot \Delta \mathbf{v}\right) V\left( \mathbf{r}-\mathbf{v}%
\right) V\left( \mathbf{r}+\mathbf{v}\right)  \nonumber \\
&=&\frac{1}{2}\int d\mathbf{x}\int d\mathbf{y}\exp \left[ i\varepsilon
\left( \mathbf{x-y}\right) \cdot \Delta \mathbf{v}\right] V\left( \mathbf{x}%
\right) V\left( \mathbf{y}\right)  \nonumber \\
&=&\frac{1}{2}\left| \int d\mathbf{x}\exp \left( i\varepsilon \mathbf{x}%
\cdot \Delta \mathbf{v}\right) V\left( \mathbf{x}\right) \right| ^{2}.
\label{F30}
\end{eqnarray}

After including the dimensional quantities $
\rlap{\protect\rule[1.1ex]{.325em}{.1ex}}h%
$ and $m$, we get the probability 
\begin{equation}
P(\mathbf{r}_{0},0\mid \mathbf{r},t)\varpropto \left| \int d\mathbf{x}\exp
\left( i\mathbf{x}\cdot \Delta \mathbf{k}\right) V\left( \mathbf{x}\right)
\right| ^{2},\Delta \mathbf{k}\equiv \frac{m}{
\rlap{\protect\rule[1.1ex]{.325em}{.1ex}}h%
}\Delta \mathbf{v.}  \label{F20}
\end{equation}
The result of the calculation agrees with the standard Born approximation
for the scattering of a particle by a potential. However the interpretation
is here quite different. The usual derivation is made in terms of waves and
after that it is necessary to make an additional assumption, quite strange
for waves, namely \emph{to interpret the amplitude squared as a probability}
(this is the celebrated Born interpretation of the wavefunction). In sharp
contrast, the obvious interpretation in our derivation is that a particle
travels with constant velocity from the source to the target, it is
deflected by the potential and again it travels with constant velocity to
the detector. The expression eq.$\left( \ref{F20}\right) $ gives precisely
the (positive) probability of a deflection with the change of velocity $%
\Delta \mathbf{v=v}_{out}-\mathbf{v}_{in.}$ It is not necessary to pass from
a wave to a particle picture by introducing the probabilistic (Born)
interpretation of a wavefunction as an additional hypothesis.

\section{Discussion}

The interpretation of the Feynman path integrals formulation of
(non-relativistic) quantum mechanics as giving the probabilities of the
different paths of a particle in going from a space-time point to another
one is appealing, but it presents at least three difficulties.

Firstly it is not guaranteed that, after integrating in the variables $%
\mathbf{v}_{j},$ the transition probability, eq.$\left( \ref{F15}\right) ,$
becomes a sum of (\emph{positive}) probabilities of the different paths,
defined by the points $\left\{ \mathbf{r}_{0},\mathbf{r}_{1},...\mathbf{r}%
_{n-1},\mathbf{r}\right\} .$ I have shown that this is the case for the free
motion and for the motion in potentials quadratic in the coordinates, but it
may be not true for other potentials. However it might hold true for all
physical (sufficiently smooth) potentials, but this is a speculation. This
problem is not soved in this paper.

Secondly the probabilistic and nonlocal action of the potential on the
particle shown in eq.$\left( \ref{F30}\right) $ is certainly strange. It
strongly departs from the classical (deterministic and local) action
governed by Newton\'{}s second law. In relation with this objection I may
argue that, although strange, the action is compatible with the ``ontology''
of classical physics, that is particles are (localized) corpuscles. The
nonlocal character of the action might be due to some random motion
superimposed to the smooth path, hidden in the effective law of force, eq.$%
\left( \ref{F20}\right) $. On the other hand non-relativistic quantum
mechanics is an approximation of a relativistic field theory, whence the
formalism here studied may be also an approximation to a more accurate
theory.

Thirdly the probability of the travel from one point of space-time to
another one does not cover all the richness of quantum mechanics, in
particular it does not contain the Heisenberg uncertainty relations.

In summary it is not obvious whether the development of Feynman\'{}s
formalism here presented possesses a physical interest or it is just a
mathematical exercise devoid of any physical meaning. In any case it
illustrates the fact that a direct interpretation of the quantum formalism
may not be the only one possible. For instance the interpretation of Feynman
path integrals in quantum mechanics should not mean that ``the particle goes
via all possible paths simultaneously'' as sometimes stated.

\end{document}